\documentclass[preprint]{revtex4} 
\usepackage{graphicx}
\usepackage{setspace}
\usepackage[T1]{fontenc}
\usepackage{amsmath}
\usepackage{amssymb}
\usepackage{amsfonts}
\usepackage{amsthm}
\usepackage{multirow}

\begin{document}

\author{Marco~Cosentino Lagomarsino}
\email[e-mail: ] {Marco.Cosentino-Lagomarsino@unimi.it}
\author{Bruno~Bassetti}
\affiliation{Universit\`a degli Studi di Milano, Dip.  Fisica. Via
  Celoria 16, 20133 Milano, Italy} \altaffiliation{and
  I.N.F.N. Milano, Italy.}

\author{Gastone~Castellani}
\author{Daniel~Remondini} \email[e-mail: ] {daniel.remondini@unibo.it}
\email[fax: ] {+39-051-2095050} \affiliation{Universit\'a di Bologna,
  Dip. Fisica. Viale Berti Pichat 6/2, 40127 Bologna, Italy}
\altaffiliation{and I.N.F.N. and C.I.G. Bologna, Italy.}

\title{Functional Models for Large-scale Gene Regulation
  Networks:\\Realism and Fiction}

\begin{abstract}
  High-throughput experiments are shedding light on the topology of
  large regulatory networks and at the same time their functional
  states, namely the states of activation of the nodes (for example
  transcript or protein levels) in different conditions, times,
  environments. We now possess a certain amount of information about
  these two levels of description, stored in libraries, databases,
  ontologies. A current challenge is to bridge the gap between
  topology and function, i.e. developing quantitative models aiming at
  characterizing the expression patterns of large sets of genes.
  However, approaches that work well for small networks, become
  impossible to master at large scales, mainly because parameters
  proliferate. In this review we discuss the state of the art of
  large-scale functional network models, addressing the issue of what
  can be considered as ``realistic'' and which the main limitations
  can be. We also show some directions for future work, trying to set
  the goals that future models should try to achieve. Finally, we will
  emphasize the possible benefits in the understanding of biological
  mechanisms underlying complex multifactorial diseases, and in the
  development of novel strategies for the description and the
  treatment of such pathologies.
\end{abstract}


\maketitle

\section*{Introduction}

Cells are able to respond to environmental changes and variations
of their internal state (exogenous and endogenous stimuli) through
the coordinated and regulated expression of large sets of genes.
Understanding this collective behavior is an important biological
challenge.  In general, regulation of gene expression is exerted
through several steps, the first of which is transcription into
mRNA, modulated by the combinatorial binding of transcription
factors and adapter proteins.  This step is followed by possibly
several post-transcriptional regulation
events~\cite{Boyd08,Mello07,Guarnieri08}, eventually leading to
translation into a protein, which can in turn directly or
indirectly regulate transcription. Since the number of genes and
of gene functions are fairly high, and genes may interact in
complex ways, a useful approach to understanding the behavior of
cells is to characterize this system from a global, or "network"
point of view.

\begin{figure}[phtb] 
  \centering
  \includegraphics[width=1\textwidth]{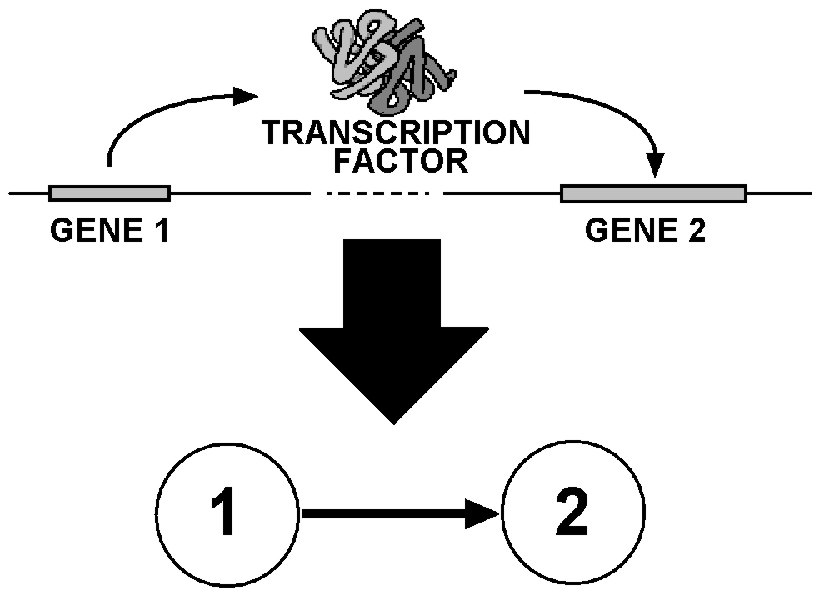}
\caption{Illustration of a network-based model.  The elements of the
  modelled system (genes and/or proteins) are represented as nodes, and
  the existence of a relationship between elements (e.g. inhibition by
  one gene transcript on the production of another gene transcript by
  means of a protein/DNA binding) is represented by a link between
  nodes. Links may be directed or not, that is relations can in
  general be asymmetric ("1" influences "2" but the opposite is not
  true). The structure of the relationships (network topology)
  determines the main system properties at a global scale, also
  imposing a hierarchy at the single node level. In ``functional''
  models, variables are defined on the nodes, representing e.g. gene
  expression. }
  \label{fig:1}
\end{figure}

Since transcriptional regulation is one of the basal cellular
mechanisms by which cells perform the variety of responses to
stimuli, through simple interactions or complex pathways
\cite{RS93,PWG99,PC00,BB04,UKZ05,BBA07}, it is possible to
simplify the scenario by considering this kind of interaction
alone, as a direct action of a transcription factor (TF) on a
target (a gene or another TF).  To this aim, transcriptional
regulation networks are defined starting from the basic functional
elements of transcription~\cite{BLA+04}.  This information is
represented as a \emph{directed graph}, (Fig.~\ref{fig:1}) to
which we will refer to as a Transcriptional Network (TN), usually
identifying each gene transcript and their protein products with
an unique node, and each regulatory interaction with a directed
edge \( A \rightarrow B \) between the node \(B\) (the target) and
the node \(A\) (the gene coding for a transcription factor that
has at least one binding site in the cis-regulatory region of
\(B\)). With this definition, the TN structure can be obtained
both by large-scale or collections of small-scale
experiments~\cite{SMM+02,SSG+06,LRR+02,HGL+04} and its main
structural features (i.e its \emph{topology}) can be studied both
to investigate its biological implications, or to gain deeper
insight into the main evolutionary mechanisms that may have shaped
it.

In this review, we will show how in recent times the
bioinformatic/biophysic community is gradually moving from ''simple''
analysis of TN topological architecture to studies in which this
knowledge is integrated with ``functional'' considerations, concerning
the consequences of transcriptional architecture on gene expression
patterns under several experimental conditions.
Given the actual achievements of "omics" technologies, it will be
important in the near future to develop realistic large-scale
models of transcriptional regulation.  We will review the state of
the art in this field, mainly considering discrete models defined
on graphs, borrowed from Statistical Physics, and how these models
should (and could) be adapted to biologically relevant questions.
Our main goal is to discuss how such approaches, given the network
topology of the system, may be used to predict its qualitative
functional behavior, even in variable environmental conditions
that are well known to affect gene expression
patterns~\cite{PO05,MO07}.

In order to connect TN topology with gene expression, it is necessary
to develop models that must be simple enough to be computationally
(and possibly analytically) tractable, but at the same time they must
grasp the essential features of biological complexity.  An important
requirement for such models should be to isolate biological questions
from complex available data, and include just the most important
features to characterize the specific question addressed.
In our opinion, the models of Statistical Physics that are currently
used for these studies lack some important features, but they may be
useful starting points nonetheless, given their flexibility, relative
ease of treatment, and richness in behavior.

We will restrict our discussion to networks defined by
physical/chemical interactions (transcription, metabolic, or
protein-protein interaction networks) directly obtained by
experiments (like Chip-CHIP or two-hybrid techniques
\cite{Li04,Horak02}) or by suitable databases representing
up-to-date biological knowledge (like KEGG
\url{www.genome.ad.jp/kegg}, Panther \url{www.pantherdb.org},
Reactome \url{www.reactome.org}, Mint
\url{http://mint.bio.uniroma2.it/mint}). Though the ultimate goal
is to model these heterogeneous data sets, their complexity puts
them far from the ideal starting point for model development. It
is preferable to build up from simple or simplified descriptions,
by adding up ingredients gradually~\cite{GJP+08}.

We will not take into account the problem of network inference or
reconstruction starting from indirect measurements (see
\cite{Bans07,Cho07,NRF04,Rem05,Gard05} as references of network
reconstruction methods starting from different experimental designs of
gene expression measurements).

The first section of this review focuses on the topology of TNs,
from the concepts of ``network motifs'', to the global features
such as e.g. the hierarchical structure of TNs. Subsequently, the
problem of the evolution of TN topology is addressed.  The section
ends with describing studies that consider the direct relations of
TN topology with gene expression and cell response to different
stimuli.  In the second section, functional models are introduced
in which agents (genes) are represented by the network nodes and
interact on the basis of TN topology.  The main mathematical
models used so far are presented: we survey their principal
features and the main constraints to be accounted for in order to
be biologically sound.  In the final section we resume the ideas
presented, giving indications for the future directions that may
be undertaken, and showing some very recent examples that in our
opinion are moving in the "right direction".

\section*{Function and Topology}

In this section, we review the current approaches to
(transcription) network analysis based on the study of topological
features.  We will describe the current analyses of the empirical
data, starting from the pure set of all possible interactions and
gradually adding ``functional'' information regarding nodes and
their interaction.

\subsubsection*{Topology Alone}

At the lowest level of description, the main functional features
of the architecture of TNs can be inferred by studying their
topology. Topological analysis is able to capture functional
properties, and important architectural features of the
network~\cite{MIK+04,WtW04,TB04,MBZ04,MKD+04,YG06,CLJ+07}.
The most basic example of this kind of observable is the degree
sequence (number of incoming/outgoing edges) of each
node~\cite{IMK+03}. In particular, degree sequences that follow a
scale-free distribution are widespread in biological networks of
different kinds, and have been connected to a variety of
functional aspects~\cite{BO04}.

An example of topological features that have revealed useful to
gain biological insight are the so-called \emph{network
motifs}~\cite{Alo07,SMM+02}, small \emph{subgraphs} (i. e. subsets
of the whole TN) that appear with significantly higher frequency
in empirical networks as compared to suitably randomized network
realizations~\cite{MSI+02,FBJ+07}.
Because of their small size within the network (less than 10
nodes), the dynamics of these circuits can be understood with
quantitative models, such as ordinary or stochastic differential
equations~\cite{Alo07}. Experimental studies targeted on specific
network motifs~\cite{Alo07,REA02,MA03} have confirmed the
importance of these sub-circuits, and the fact that in many cases
they can work independently from the interactions surrounding
them. On the other hand, parallel studies have pointed out that in
some cases a well-defined functional identity of network motifs is
difficult to find~\cite{MBV05}, so that the discovery of these
topologically relevant subgraphs should not lead directly to their
identification as functional sub-circuits. For example, in some
cases the subgraph nodes are found to be active in very different
conditions.  Moreover, on abstract grounds, it is known that
subgraphs interact topologically with the graph's large-scale
organization~\cite{VDS+04}, so that their presence might reflect
an higher level of organization.  Overall, despite network motifs
are a very powerful concept, it seems plausible that looking at
them as independent building blocks might be an
oversimplification, and it is more realistic to regard them as
functional modules that are deeply wired into the global structure
of the network.  Thus, even a full understanding of their
functionality will likely not exhaust the subject of large-scale
transcriptional regulation. Consequently, while the challenge of
constructing functional models using topology and other biological
knowledge has been tackled in different
ways~\cite{WA03,MBZ04,RSM+02}, large-scale functional models for
gene expression do not cease to be needed in order to understand
the behavior of larger modules or entire biological networks.

Two important and related large-scale properties of the known TNs
are the hierarchical feedforward layered structure of
TFs~\cite{MBZ04,YG06,CLJ+07}, and the feedback, which in bacteria
is often essentially limited to a rather large set of
autoregulations~\cite{THP+98,CLJ+07}. The existing pyramidal
hierarchical structures, with most TFs at the bottom levels and
only a few master TFs on top, have evoked the ``organization
chart'' structures observed in social networks.
Notably, while the distributions of small network features (such
as three-nodes network motifs) seem to be common among known
transcription networks of different organisms (e.g. \emph{E.~coli}
and yeast)~\cite{MIK+04}, global topological observables,
long-range feedback among several TN layers~\cite{JB08}, hierarchy
in the interactions~\cite{Sel08} and co-regulatory
structure~\cite{BBA07} seem related to increasing organization
complexity, in parallel with the increase in complexity in
unicellular organisms, e.g. from \emph{E.~coli} to yeast.

While the relevance of ``pure'' network topology is clear from the
studies presented so far, it would be naive to believe that
network structure strictly determines the dynamics of a network.
Two basic reasons for this are that, at fixed topology, signal
integration functions may have large combinatorics~\cite{BGH03},
and at fixed topology \emph{and} signal integration functions, the
dynamics depends on the structure of inputs.  This is evident for
example at the level of network motifs, where the richness of
possible behaviors at fixed topology has been elegantly proven
using ordinary differential equation models~\cite{Ing2006}. In
order to constrain the dynamics to a specific behavior, it is
necessary to supplement topological information with a specific
model for the promoters~\cite{BBG+05} and for the kinetic
parameters, and dynamic time series experimental data should be
available for model validation.

We remark that this sort of data are generally not available or
highly stylized for large-scale analysis.  For example, the
best-known electronically encoded database, that of
\emph{E.~coli}~\cite{GJP+08}, represents signal integration as
annotations to the activity of a transcription factor at a given
promoter (activator, repressor od dual), and defines the input
structure by the nodes that are active in different growth
conditions.
In general, all the parameters necessary for developing a detailed
model are not available, thus a request for an \emph{optimal}
model is to be \emph{robust} to such lack of fine details. It
should rely only on a limited number of parameters, that should be
obtained as a sort of "principal components" of the available
experimental evidences.

\subsubsection*{Topology And Evolution}

The functional role of topology can be further investigated by
taking an evolutionary point of view. The idea is that if we
understand the ``target'' topological features of a network during
evolution, this will give us information on its functioning.
Evolution of a transcription network is driven by three main
biological mechanisms: (\textbf{i}) gene duplication,
(\textbf{ii}) rewiring of edges by mutation/selection of TF/DNA
interactions and (\textbf{iii}) horizontal gene transfer, which is
especially relevant for bacteria~\cite{PPL05}. These mechanisms,
in particular the first one, have been shown to play a substantial
role in TN evolution, although the extent to which they can shape
the network is
debated~\cite{BLA+04,TB04,CW03,DMA05,MBV05,LAC+08,PDA08,LP08}.
Looking at the elementary duplicated topological structures
conserved by selective pressure can give indications concerning
their functional importance. For example, network motifs have been
reported not to emerge from divergent
evolution~\cite{MBV05,BLA+04,CW03}, leading to a substantial
debate on their nature and their relationship to the rest of the
network, along the lines that we mentioned above.
On the other hand, the properties of hierarchy and feedback in the
\emph{E.~coli} network are compatible with evolution by
duplication/divergence~\cite{CLJ+07}, in which gene duplication
preserves the abundant self-regulations and the ``shallow'' layered
organization, which can be hypothesized to facilitate fast signal
propagation with short regulatory cascades, thereby optimizing the
time constraints for the expression of target genes in response to
 stimuli~\cite{MBZ04,YG06,CLJ+07}.
While TFs were sometimes acquired together with the adjacent
operon that they regulate~\cite{PDA08}, the inclusion of
horizontally transferred genes is estimated to be very slow and
based on the recruitment of existing transcription factors of the
host, as reflected by fast evolution of their cis-regulatory
sequences, and by the fact that genes resulting from increasingly
ancient transfer events show an increasing number of
transcriptional regulators.

Particularly interesting results come from recent studies that attempt
to use the knowledge of many sequenced genomes.
Lozada-Chavez and collaborators~\cite{LJC06} studied the
conservation of \emph{E. coli} and \emph{B. subtilis} TN structure
on a large number of known bacterial genomes. They found large
plasticity of regulatory interactions, witnessed for example by
the underrepresentation of significantly conserved transcriptional
regulatory interactions among different phyla of bacteria, and by
the lack of constraints on the co-evolution of combinatorial
interactions. Their results indicate that transcriptional
regulation is more flexible than the genetic component of the
organisms, and its complexity and structure plays an important
role in the process of phenotypic adaptation. Analogous results
were obtained by Babu and coworkers~\cite{MBT06}, who noted that
different transcription factors have emerged independently as
dominant regulatory hubs in 175 prokaryotes, suggesting that they
have convergently acquired the observed scale-free (or at least
heavy-tailed) out-degree sequence.  Also, common lifestyle brings
to conservation of orthologous interactions and network motifs
over a large phylogenetic distance, indicating that network
structures are conserved if the selective pressure of
environmental stimuli is present.

A critical point in these studies is that the network topology is
known to an acceptable extent for very few TNs only (essentially
the \emph{E. coli} and \emph{B. subtilis} sets used in the two
studies) so that arbitrary, though reasonable, projection
algorithms have to be used in order to infer the network topology
of genomes with known sequence, creating the danger of circular
arguments.

\subsubsection*{Topology Studies Including Functional Features}

A further step towards functional models that relate networks to
gene expression is the bioinformatic integration of data
concerning the activity and roles of promoters, targets and TFs.
The most important step in this direction is probably the work by
Luscombe and coworkers~\cite{LBY+04}, who integrated, in a pioneering
study on yeast, transcriptional regulatory information and
gene-expression data in multiple conditions.  These authors have
observed that not all network interactions are always accessible to
the system, but only specific network subsets are switched on
simultaneously. Accordingly, only few transcription factors serve as
permanent hubs, while most act transiently and are active only under
certain stimuli. Thus the effective interaction topology, i.e. the
network formed by the transcription factors that are active at the
same time in response to a specific perturbation, varies greatly in
different conditions and environments, and may differ substantially
from the global interaction topology.

This point has broad implications. For example, from the point of
view of inference, it implies that models constructed based on
data obtained under different conditions or even from different
organisms may be strongly deceptive since they include
interactions that do not occur at the same time~\cite{Nord07}.
This point must be taken into account when considering the network
structure to be used for modelling (discussed in the following
section).

By studying these sub-networks, they show that environmental
responses facilitate short-time signal propagation, whereas
processes driven by internal stimuli such as cell cycle and
sporulation proceed by successive slower stages (see \cite{Bahl08}
for a recent review on these aspects of transcriptional
regulation). The former signals are also overrepresented for
direct transcriptional feedback~\cite{JB08}.
Another study by Janga and coworkers~\cite{JSM+07} classifies TFs
into external-, internal- and hybrid-sensing, and shows how
different network motifs use different sensing structures, which
are also related to regulatory modes (activation, repression, or
dual regulation)~\cite{BBA07}. Finally, one can consider the
co-regulatory network of TFs sharing common targets~\cite{BBI+06},
which enables to distinguish between regulators that integrate
different cellular processes and those acting specifically on one
or a few major processes.

The above mentioned approaches are powerful, but at the same time
they are limited by a purely data-driven approach. A model-guided
analysis is potentially more powerful, because it accounts for all
the possible phenomena given certain initial assumptions motivated
by the system. As a positive example we would like to mention the
case of flux balance analysis for metabolic
networks~\cite{JP07,JP08}, a constraint-based modelling approach
that incorporates a simplified reaction stoichiometry in terms of
maximization of fluxes. This technique simplifies enough the
system making it treatable as compared to the unreachable full
stoichiometric problem, and at the same time allowing nontrivial
and correct predictions on mutant viability and phenotypes. The
key feature of this strategy is that it does not attempt to
predict the exact network behavior, but rather it uses known
physicochemical constraints to separate the states that a system
can achieve from those that it can not. Analogous efforts for
protein-protein interaction networks are being discussed in the
literature~\cite{MI07}. Such an effective level of description is
lacking for TNs.  Therefore the next step, and challenge, is to
perform integrative bioinformatic analysis directly using models
to guide them, and conversely to test models with data, in order
to find the most suitable \emph{``coarse-grained''} level of
description, as we will outline in the next section.

\section*{Hypotheses and constraints for effective functional models}

The remainder of this review will address the problem of the
possible abstract models that can describe empirical data,
considering both the network interactions and the activation state
of the nodes.  We will review some of the available candidates for
a network model that include agents (the genes) interacting
between themselves (i.e. binding to promoters and
enhancing/inhibiting transcription) through the network
connections, trying to elucidate the questions that can be posed
using these tools, and the main needs in their formulation (see
Box 1, 2). As anticipated, we believe that the most relevant needs
are conceptual simplicity and computational manageability, and
thus we will not consider models of TNs that take into account the
fine details of (bio)chemical reactions, which immediately lead to
a proliferation of unknown parameters (e.g. reaction
constants)~\cite{HKHR08,VrL06}. While these models are well
established for small regulatory circuits, it is not feasible to
simply ``scale them up'' and model global expression states.
Rather, we concentrate on effective \emph{coarse-grained} models
which reduce to the minimum the number of relevant parameters.
It should be obvious, given the previous discussion, that such
models cannot aspire to describe a specific TN in detail. Rather,
they should be able to characterize (in a statistical fashion) the
behaviour of \emph{classes} of networks or subnetworks with
similar topological or functional structures.  We will also limit
the discussion to models in which only discrete states are allowed
for each transcript concentration (or level of gene activation),
in contrast with models in which continuous values are considered
(like those based on Mass Action Law, such as Michaelis Menten
kinetics). We do not exclude that continuous variable models can
be useful, but we believe that the assumptions of treatability and
low number of effective parameters should hold also in this case.

\subsubsection*{Dynamics versus Optimization}

Most of the currently available coarse-grained models focus on
dynamical features. The prototype of these models are Random
Boolean Networks (RBN), which have a long history starting from
the original Kauffman model of the late
'60s~\cite{Kau69,Kau93,Ger02}. RBNs have been studied extensively
on theoretical grounds since the
'80s~\cite{DerPom86,CKZa01,CKZb01,GreDro05}, and have recently
received renewed attention by the introduction of new connectivity
topologies~\cite{Ald03,MiDro06}, that led to new results in terms
of possible dynamic states of the network. RBNs are typically
studied under a wide set of random initial conditions, allowing to
characterize the basins of attraction of the final states of the
system.

The classic Kauffman model is formulated to capture on abstract
grounds the "dynamic" behaviour of the cell: N two-state genes
(on-off, 0-1, true-false as in Boolean logic) change state as a
function of the states of their regulators.  In the original
model, the network ensemble is constrained to K random regulators
per gene, and the control functions are chosen randomly among all
possible Boolean functions.  Depending on the initial conditions,
several \emph{attractors} like fixed points (i.e. a fixed state
for each element of the system) or cycles (in which a limited
number of system states are periodically visited) may exist, with
basins of attraction of different size.  These different
attractors of the system are commonly interpreted as specific
biological tasks performed by the cell, or as specific cell types
(phenotypes) arising from a differentiation process, and the
system is characterized on the basis of the number of these
attractors, their typical length, and the probability of switching
among them due to noise or external perturbations. In the case of
"classical" RBNs, in which all nodes update synchronously, the
possible phenomenology includes fixed points or cycles of length
at most $2^N$.

Also, a number of variants of this model are now available
\cite{Schmul03,Ger02}. Each variant of the model can be further
characterized by 1) the topology of the network, 2) the choice of
the Boolean functions (that may differ among nodes) and 3) the
update rule.
Varying K (the inward connectivity degree), the system may undergo
a transition between a state of ordered dynamics (in which
perturbations to the system tend to die out) to one of chaotic
dynamics, characterized by large cycles (of exponential length in
$N$) in which even a small perturbation may propagate throughout
the whole system indefinitely.
The study of possible dynamic states as a function of more complex and
biologically relevant topologies is developing in the latest years
\cite{Ald03,Schmul03,HKHR08}, in which the relevant feature of the
system are controlled by a limited number of parameters (e.g. the
exponent of the power-law distribution of node connectivities).
Also the set of Boolean functions, chosen to represent gene
activation/inhibition, can influence system dynamical transition: for
example, the choice of more ``canalizing'' functions (i.e. less
variable with respect to flips in the state of a small number of
inputs) can control the length of the typical cycles
\cite{Ger02,Schmul03,Niko07}.
Regarding the update rules of the network, several options are
available \cite{Ger02}: synchronous update, in which all nodes are
updated together; random asynchronous update, in which one or more
nodes are chosen randomly at each step, and cyclic update, in
which one or more nodes are updated at fixed periods. These update
rules share common properties (e.g. the number of fixed
points~\cite{LJB05,CLP06}), but also show a different behavior in
the size of the basins of attraction and in the existence and size
of cycles. For random asynchronous update rules there cannot be
cycles, but network states may get trapped into \emph{loose
attractors} in which the same group of states may be visited
repeatedly but with a different order. This property essentially
puts random-update asynchronous models in relation to
``equilibrium'' models, in which it is sufficient to supply the
behavior of the attractors for a complete characterization (see
below).
Even if the completely synchronous updating rule may be
unrealistic for a real TN, a certain degree of coordination in TN
activation must surely be present. The interplay between cyclic
updating (resembling \emph{housekeeping} tasks like cell cycle or
circadian rhythms) and random update (due to response to external
stimuli) might be interesting to investigate, since it is very
likely that a real TN combines a mixture of both updating rules.

Other discrete dynamical models, that allow more complex
behaviour, are Thomas' model \cite{Thomas95,Thomas98}, and
Petri-nets \cite{Hardy04}. Thomas' model is essentially an
asynchronous-update discrete network in which also multiple values
of the state variables are allowed. The asynchronous update is
generally not thought as randomly ordered, but sometimes used to
introduce a time hierarchy.  Petri-nets are cellular automata
describing the flows of masses through a network, and thus
especially well-adapted to describe metabolic networks. However,
they also have been applied very recently to different regulatory
networks, including TNs~\cite{SBS+07,GSH+08}.

Both Thomas' model and Petri nets are formulated to introduce more
realistic features as compared to the simplest Boolean models,
prominently time and time-hierarchies related to biological
complexity~\cite{GSH+08,Thomas95}, while preserving a
coarse-graining description of the states. As such, they are
especially useful to model mid-sized networks, in which a
continuous approach (as with ordinary differential equations) may
fail, but a Boolean description might be too crude. They have been
mainly applied to inference problems from gene expression data, or
to small network models, and generally not in a large-scale
context (see for example \cite{SchBraz07} for a more extended
discussion). More systematic large-scale studies are needed in
order to establish the distinct features of their typical
behaviour.

A different approach is to formulate the model as an optimization
problem, in which the addressed question is to find the
``equilibrium'' gene expression patterns compatible with a set of
``constraints'' set by the network interactions. This formulation
is simpler than considering system dynamics, but possibly still
useful to extract relevant biological information (see Box~1). It
is more suitable for problems in which time is not a central
issue, for example studying the response of the cell to
environment or (pharmacological) treatments that may alter its
equilibrium state, as opposed to transient gene expression changes
during the cell cycle. In general, such a model can be useful in
studies concerning the structure of stable states and basins of
attraction of the system as a function of network
topology~\cite{CJB05,CLP06}, and how they may change if the
network structure is perturbed. It has also been used to
characterize the behavior of random-update Boolean
networks~\cite{MPW+07}.
The reference models for such equilibrium studies are the so
called ``spin models''.  These are one of the most "versatile"
models in Statistical Physics, and have been successfully applied
in several contexts.  Their terminology is rooted in their
original physical application, describing the discrete atomic
states (magnetic \emph{spins}) of ferromagnetic materials.
Biologically relevant results have been obtained, to cite only some of
them, in their adaptation and application for memory formation and
information storage in neurons~\cite{Hopf82,Opper87}, in the
description of stability properties in biological
membranes~\cite{Mutz91,Carraro93} and in the spread of political
opinion in social networks~\cite{Indekeu04,Castellano05}.
In the simplest spin model, only two states are allowed for each
element of the system (corresponding to active/inactive expression
states of a gene, as in RBN) but it is possible to consider an
arbitrary number of states (Potts' model~\cite{PottsWu82}), that
in case of gene expression models could mimic multiple levels of
activation (e.g. due to a different combination of transcription
factors bound to the gene promoter regions). Spin models have been
recently applied to models of gene regulation
\cite{Bar04,ZhouLipowsky05,Castellano06}, to describe how some
network features can affect global stability properties.

\subsubsection*{Choosing the interactions}

An important problem for any realistic modelling of a TN is how to
represent the biological interactions, e.g. how the signals arriving
at promoter regions are integrated for transcription initiation.  For
many organisms, the exact structure of TN has not been described yet:
while the presence or absence of an interaction is often known, its
actual role in inhibition or activation is not so often annotated in
the data sets.
Moreover, it is clear that TF regulation is a so-called many-body
interaction (involving more than two TFs at a time), since the
combinatorial action of several TFs may be paramount in determining
the output.
In this sense, the typical choice of general classes of Boolean
functions in RBNs is biologically realistic. It can be also shown,
using the Shea-Ackers' model \cite{SA85,BGH03} that signal
integration can realize all Boolean functions. Spin models as well
can have many-spin interactions.
On the other hand, it is not known which functions are effectively
realized in a biological context. For example, many Boolean functions
can be constant with respect to a number of inputs, while in a
biological situation, if an interaction exists, it is very probably
used in at least some context, so that the problem of selecting
``realistic'' functions is still open \cite{MA05}.
Large-scale data on the effective combinatorial functions involved
in transcriptional process are largely unavailable, except for few
well characterized promoters. To date, the work of Davidson and
coworkers is the most complete in characterizing the
transcriptional interaction logic for a large number of network
interactions in developmental pathways \cite{ID05,MD07}.
For some \emph{E.coli} \cite{SGPC06} and \emph{Yeast} TN data
sets, annotations are available concerning the activating or
inhibiting role of TFs participating in transcriptional
interactions \cite{MasSnep05,MSI+02}, so it would be possible to
consider classes of functions compatible with these constraints,
or to simplify the model using pairwise interactions based on
these annotations.

Given the heterogeneity of interactions, the most suitable spin
models for this purpose may fall into a subgroup referred to as
\emph{spin glasses}~\cite{MezParVir87}, in which spin-spin
interactions vary along the network and may give rise to the
phenomenon of \emph{frustration}, i.e. the fact that not all
cooperative/competitive relationships can be satisfied at the same
time, leading to multiple equivalent clusters of compatible
states. These models have been widely characterized theoretically,
also leading to the development of analytical solution tools
("replica" and "cavity" methods, see~\cite{MezParVir87,MPZ02}).
Note that, however, if connections are not symmetric (as in real
TNs), the scenario can be much more complex as compared to the
results available with spin-glass models.

\subsubsection*{Which topology for a plausible model?}

As emerging from previous discussions, the first key point in
obtaining biologically plausible results is thus related to the
introduction of an adequate network topology.
If we consider, for example, the fixed in-degree ensemble of classical
Kauffman networks, this topology is clearly unrealistic as compared to
known TNs.  First, there are no input nodes (i.e. only aimed at
receiving signals), but all nodes equally receive and send signals.
Second, the number of incoming connections is the same for each node,
while empirical TNs have been observed to have different incoming and
outgoing degree distributions, with the second group much more
heterogeneous than the first \cite{GBB+02,BO04}.
Thus, to make the above mentioned models directly applicable to
empirical networks, they must be constrained as much as possible by
all the biological knowledge available from experimental data,
including some of the empirical features discussed above, such as
network motifs, hierarchical structure of nodes, degree correlations,
and loops.

In many network-based models, topology is crucial in altering
equilibrium as well as dynamical properties of a system
\cite{Strogatz98,Vesp01,Lat06}.  Some recent
investigations~\cite{Ald03,Bar04,ZhouLipowsky05,Castellano06} have
introduced more realistic degree distributions for network
connectivity in RBN and spin models, showing how some properties
of the system depend critically from the exponent of the power-law
connectivity degree distribution.  The scale-free property of
connectivity degree distributions, as observed in many biological
networks~\cite{BO04,RSM+02} is only a first step: many other
structures can be embedded in a scale-free network (and in any
other network topology as well) such as for example
assortativity/dissortativity \cite{MasSnep02,Newman02} or non
trivial relationships between network observables (e.g. centrality
measures~\cite{Rem07}) or the presence of feedback loops
\cite{JB08,CLJ+07}.  The role of these topological features is
still to be investigated. As an example, recent experimental
results \cite{Janga08} show the existence of a "metrics" on
chromosomes, such that genes activated by the same TF tend to be
closer in terms of base-pair distances. The same kind of effect
can be caused by chromatin remodelling, which introduces
correlations in the activation of close genes, even if they are
not coregulated~\cite{Eis06,Hurst07}. These observations suggest
the existence of correlations in TN connectivity, facilitating the
coordinated expression of genes with similar tasks.

\subsubsection*{Introducing environment and perturbations}

In a cell, the signalling machinery transfers to and integrates in
the regulatory networks a rich variety of \emph{external} stimuli,
such as changes in nutrient concentration, hormonal signals or
pathogenic aggressions. This corresponds to introducing external
constraints in the model, or ``fields'', which may influence the
activation states of specific groups of nodes. For this purpose, a
realistic model should include in the first place a set of
receptor or input nodes, and it should account for the possibility
of having specific subnetworks activated (or silenced) by proper
input patterns. As stated previously, we believe this feature is
biologically relevant, since depending on the state of the cell
only some subnetworks may be effective~\cite{LBY+04}, rather than
having the whole set of possible interactions activated at every
time.
Moreover, it is not unusual that several signals reach the system
concomitantly and repeatedly (e.g. during differentiation or
apoptosis) forming specific input patterns.  Ideally, a model
could allow \emph{in silico} experiments in order to reproduce
qualitatively experimental results, or to verify system properties
like robustness to changes in the input pattern (as it has been
done for the case of the yeast cell cycle~\cite{LiTang04,OLB+08}).
The effect of varying input structures has been recently addressed
experimentally with a systematic approach applied to a discrete
Boolean model. In \cite{HKHR08}, the regulatory machinery,
interpreted as an "information processing network", has been
subjected to a large number of random input combinations, and the
analysis of system response has revealed its capability to perform
sharp classification tasks even in presence of added noise.

One of the simplest questions along these lines could be to study the
effect of a perturbation to a single network element. In RBNs, the
results of single-site perturbation have been studied by allowing a
node to switch its state arbitrarily and observing the effect on
system dynamics (e.g. the possibility of a transition to a different
system state). However, such an approach is not feasible for modelling
the outcome of recent experiments, in which a single gene is
conditionally activated or inhibited (see \cite{Sed03,Rem05}), and
only recent studies~\cite{FreDro08} are beginning to apply
perturbations to RBNs that may resemble biologically plausible
situations.
This is analogous to the case of organisms with fully sequenced
genomes, in which systematic methods have been used to analyze,
both from a proteomic and genomic point of view, the effect of
suitable perturbations on specific pathways (as in \cite{IH01}).
Variations on the theme of ''attack-tolerance
models''~\cite{AJB00,Tieri05} have shown that single network
elements can influence the system properties to a different
extent, revealing a complex hierarchy based on system topological
structure. It is surely an important question to check if the most
relevant elements in a network from a purely topological point of
view (in the sense of a \emph{static} description of the network)
are also the most relevant in influencing the \emph{dynamical}
properties of the system. In facts, network dynamics following a
specific perturbation could also depend on other factors, such as
the interaction rules (e.g. the activation functions) chosen at
each node.

More complex ''ideal experiments'' can be performed, by
considering the concomitant switching of groups of nodes chosen on
the basis of their role inside the network: e.g. how a stimulus
can be propagated through the network acting on ''core'' nodes as
compared to ''peripheral'' nodes, ranked by different centrality
measures. Such \emph{in-silico} experiments may help in reducing
the number of real experiments to be performed, that in general
would ''explode exponentially'' with the number of genes to be
perturbed together, evidencing the potentially most interesting
situations. Viceversa, different network-based ranking of the
nodes could be checked experimentally, providing deeper insight
into the analogies between models and real TNs.

\section*{Discussion and Conclusions}

In this review, we have reported the increasing need of
quantitative models for the interpretation of large-scale
experiments on gene regulatory networks, that are continuously
increasing in number, accuracy and quality of the generated data.
We surveyed some of the available models for such purposes. The
landscape comprises a wealth of possibilities, but a unified view
is still missing, mainly due to the fact that most available
models can only be studied on a very abstract level, and the
possibility of a comparison between experiments and models is
opening only in recent years.

The key issues arising in this field that we aimed to address are
the following. First, not all global properties of networks may be
simple consequences of the behavior of small sub-parts or
sub-networks, so that there is the need of models able to deal
with the regulatory network as a whole. Secondly, models that are
very useful for small-scale systems generally fail when scaled up,
because of parameter proliferation: a large-scale model must meet
the requirement of being based on a small number of relevant
degrees of freedom. Thirdly, as a consequence of the two previous
statements, such large-scale models cannot be all-inclusive, but
must be targeted to specific biological questions. The key
ingredients for the model should then be isolated from the
plethora of available data, selecting just the most important ones
to characterize the specific question addressed.

Accordingly, we have proposed some constraints that TN models
should satisfy in order to be ``realistic'', and some possible
questions that could be isolated.  A first key point is simplicity
and treatability (i.e. containing a small number of parameters to
be "tuned") but at the same time they must be able to describe the
main features derived from experimental observations, for example
regarding the interactions between nodes (network topology) and
the way a node may elaborate the incoming signals (the activation
function). In our opinion, some key features have not been taken
into account yet, regarding both the structure of trans-regulation
relationships and the detailed modelling of the environmental
conditions in which the system is embedded, in particular how the
model may receive stimuli and respond to them, which is a key
point in describing realistically a biological system.

Based on what we described as reasonable assumptions for a
realistic model, it may become possible to address some more
specific questions of biological relevance.  For example, a
characterization of the structure of system "attractors" as a
function of network topology could help answer to questions such
as whether it is possible to characterize the states of a cell as
unique (e.g. a unique "healthy" or "basal" state). This issue may
be important for practical applications, because apparently
identical cells (i. e. with similar phenotype) could respond
differently to external threats, depending on the effective
regulatory state they occupy (a very recent paper~\cite{Chouard08}
discusses this issue in detail, posing the question also in terms
of the evolutionary advantages of such ''redundancy'').

This point is emerging in recent studies on genomic profiling of
tumors\cite{MLL02}, where new tumor subgroups are found that are
not recognizable by usual phenotypic observations of cells and
tissues. This refinement of tumoral cell classification is leading
to new insights in the underlying biological mechanisms, and also
opening new directions for the choice of optimal therapeutic
strategies (e.g. in terms of pharmacoresistance or tumor
aggressiveness level).

An interesting example comes from the work of Kitano and
collaborators \cite{CseteDoyle04,Kitano04,Kitano07}, in which
topological "double funnel" structures called \emph{bow-ties} are
repeatedly found in cell signaling system at various levels. The
hypothesis is that specific elements in this structure are more
suitable as targets for therapy, mainly on the basis of their role
in the regulatory network rather than on their biological function
only.

Also the role of external perturbations affecting the system (from
single-gene knock out to the typical flow of biological information
entering a cell) and the role of other modulating factors (chromatin
structure and post-translational mechanisms) are open to more detailed
investigations with the help of suitable models.

A fundamental question, in many biological contexts, is to understand
if external stimuli produce a "perturbation of the basal state" (to
which the cell should thus be able to return to), or if they may
induce a switch to a different stable state, for example in the case
of tumors, Alzheimer disease or autoimmune pathologies.  Sufficiently
realistic models should help to clarify how many external/internal
factors, and of which entity and acting for how long, should be needed
in order to reach such "point of no return".

We conclude by noticing that recent experiments present a design
that makes them increasingly more suited for theoretical
modelling, rather than being focused on issues of direct
biological/medical relevance. This is a clear sign of the times,
reflecting the increasing interest towards novel approaches in
understanding biological processes, dictated by the new
quantitative points of view that are nowadays gathered under the
large hat of Systems Biology.  As an example, Isalan and
coworkers~\cite{ILM+08} constructed a large set of recombinations
of promoters and regulatory regions in \emph{E.~coli} (possibly
the whole set of permutations allowed for its TN) and showed that
most new networks are tolerated by the bacteria and that
expression levels correlate with factor position in the wild-type
network hierarchy. We believe that the challenge of testing a
simple and biologically realistic model in the description of
these and similar experiments on TNs is becoming feasible, and the
results will help pave the way to a new biological conceptual
framework.

\begin{acknowledgments}
D. R. wishes to thank "Progetto Strategico d'Ateneo" from Bologna
University for financial support.
\end{acknowledgments}

\bibliography{review_biblio}

\newpage

\newpage

\begin{singlespace}
\begin{center}

\fbox{
    \begin{minipage}{0.9\textwidth}

      \vspace{0.5cm}

      {\large \begin{center} \textbf{BOX~1:} Spin Models and Random Boolean Networks\end{center}
      }

     \begin{center} \includegraphics[width=0.9\textwidth]{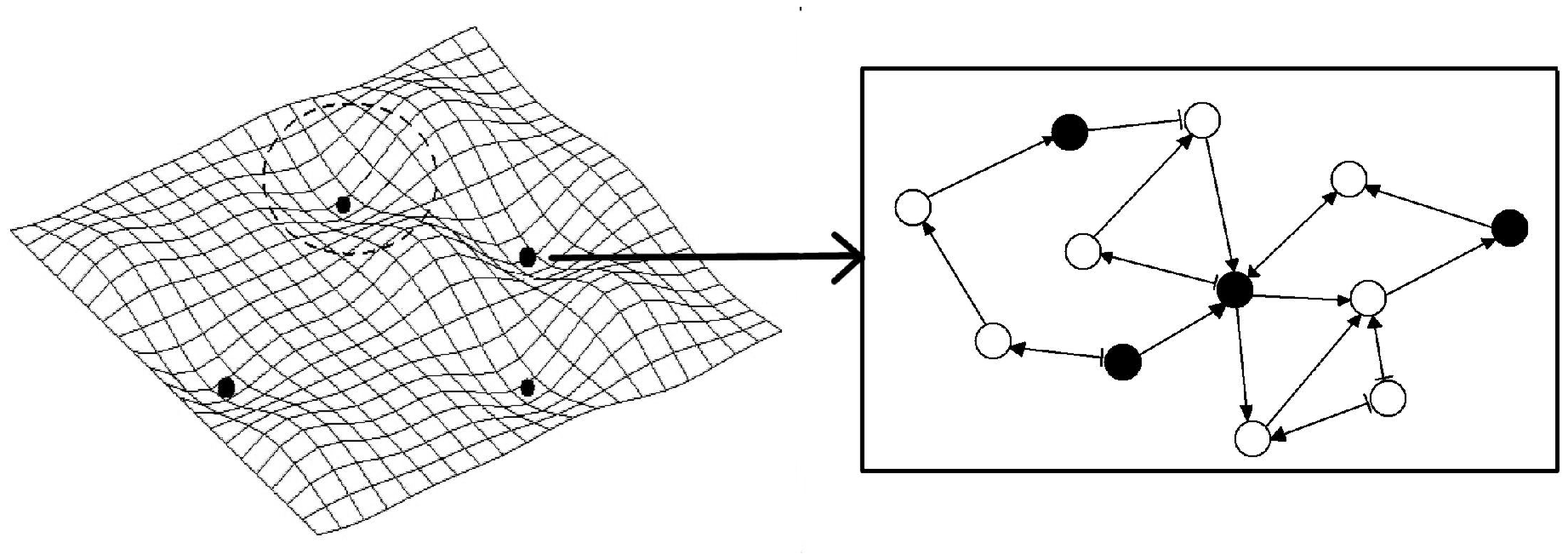} \end{center}

      \vspace{0.2cm}

      \begin{center}

        \parbox{0.9\textwidth}{In a spin model, the system has equilibrium states (right) corresponding to
          minima of an "energy function" (left); the structure of
          the energy function is defined by the regulatory
          relationships in the graph, and characterizes the stable
          state features as well as the transitions between states as
          a consequence of perturbations.

          In a Random Boolean Network, the system is characterized by the set of
          possible states and the dynamic transitions between them, with particular attention to
          fixed points, cycles (i.e. periodic dynamics) and the size of their basins of attraction.}

      \vspace{0.5cm}

     \end{center}

    \vspace{0.5cm}
 \end{minipage}

}

\end{center}
\end{singlespace}

\newpage

\begin{singlespace}
\begin{center}
\fbox{
    \begin{minipage}{0.9\textwidth}
      \vspace{0.5cm}

      {\large \begin{center} \textbf{BOX~2:} Discrete \emph{versus} Continuous
        models, a resume. \end{center} }

      \vspace{0.2cm}

      \textbf{Continuous Models} \\

      \vspace{0.2cm} {\small In \emph{kinetic models}, the network is
        described by the set of chemical dynamic rate equations
        associated to the network interactions. These models, in the
        form of ordinary or stochastic differential equations, are the
        standard description for small to medium
        networks~\cite{TCN03,MA97}. They require the detailed
        knowledge of many parameters, which is one of the main issues
        in the field~\cite{VrL06}. For this reason, it is unlikely
        that these models (with today's technology) are scalable to
        entire network descriptions. Furthermore, for the case of
        transcriptional interactions, extra models, such as the Shea
        Ackers model, are required for the description of promoter
        activity.

     \vspace{0.2cm}

     The model of \emph{Shea and Ackers} computes the probability of
     transcription initiation using a \emph{static equilibrium}
     description of binding at a known promoter, and given the binding
     affinities and interactions of all transcription factors binding
     in the \emph{cis}-regulatory region. If the mass action law is
     assumed for the products and stationarity of gene expression can
     be assumed, this model could be used to compute \emph{output
       compatible states}, given the \emph{input concentrations} of
     TFs. However, it still suffers from the parameter proliferation
     problem.  } \vspace{0.5cm}

      \textbf{Discrete Models}

      \vspace{0.2cm}

      {\small The mostly used \emph{discrete dynamic models} are
        \emph{Boolean Networks}, that have been successfully employed in
        empirical studies~\cite{OLB+08}. In this model, gene expression levels
        are simplified to on/off dynamics, and regulation is represented by Boolean
        functions. \emph{Synchronous dynamics} mimics the time
        hierarchy of events of gene regulation. These models can be
        run on ensembles of network topologies and/or activating functions,
        or specific realizations may be considered for both features.
        More complex models for discrete-state dynamics are given by
        \emph{Thomas' model}, or \emph{Petri nets}~\cite{Thomas95,Thomas98,Hardy04,SchBraz07}.

      \vspace{0.2cm}

      \emph{Spin Models} are aimed at characterizing the
      \emph{compatible states} for a set of discrete agents (the
      spins) representing the gene expression levels, interacting in
      pairs or groups through some \emph{constraints}, acting through the network topology, representing
      the \emph{cis}-regulatory interactions governing gene
      expression. These models (very abstract but flexible) can be used
      to describe the essential features of TNs and to produce
      qualitative observables related to the collective behavior
      of gene regulation in different specific problems (response to
      perturbation, robustness, inference of interactions, etc.). In
      particular, for topologies that are rich in \emph{cycles} it is possible to find
      \emph{frustration} (namely the impossibility to satisfy all constraints at the same time) and
      consequently the phenomenology of \emph{spin-glass models}, in which
      multi-scale clusters of stable states emerge.}

    \vspace{0.2cm}

    \end{minipage}
}
\end{center}
\end{singlespace}

\end{document}